\documentclass[]{article}
\usepackage{graphics} %   <------- for LatTeX + PS
\usepackage[dvips]{graphicx}
\usepackage{amsmath}
\usepackage{amsfonts} 
\usepackage{amssymb}%
\begin{document}
\title{Local excitation-lateral inhibition interaction yields wave instabilities in 
spatial systems involving finite propagation delay}
\date{\today}
\author{Axel Hutt\\Department of Physics, University of Ottawa, \\150 Louis Pasteur, Ottawa K1N 6N5, Canada\\email:ahutt@uottawa.ca}

\maketitle

\begin{abstract}
The work studies wave activity in spatial systems, which exhibit 
nonlocal spatial interactions at the presence of a finite propagation speed. We find
analytically propagation delay-induced wave instabilities for various local excitatory and lateral
 inhibitory spatial interactions. The final numerical simulation confirms the analytical results.
\end{abstract}

\section{Introduction}
In the recent decades the space-time dynamics of extended complex systems has attracted much attention in physical, chemical and biological science~\cite{HakenBook02,Nicolis+Prigogine77}. In this context the nonlocal interactions between system subunits have been studied extensively~\cite{Coombes05_2,Cross93,Wilson+Cowan72}. By virtue of this nonlocal spatial 
interaction, the propagation delay between two spatial locations may not be negligible 
if the the propagation speed in the system is small enough. Such propagation delays
have been found to affect the activity spread solids~\cite{Tzou+Chen98,Mitra_etal95} and 
plasmas~\cite{Lazzaro+Wilhelmsson98}, they affect the stability of neuronal systems~\cite{Hutt_Network03} 
or may be responsible for subballistic transport in solids~\cite{Metzler+Nonnenmacher98}. The present 
work addresses these phenomena and studies
the effect of a finite propagation speed in a generic spatio-temporal model.\\
To describe the space-time evolution of nonlocal systems mathematically, partial differential equations have been studied vastly, e.g.~\cite{Haken04book,Nicolis+Prigogine77}. Moreover integral-differential 
equations also have been studied in recent years to describe nonlocal interaction in spatially
extended neuronal systems~\cite{Hutt+Atay06,Coombes05_2,Blomquist05,Bressloff99,Wilson+Cowan72}. 
Both latter equation types are strongly related for some specific integral-differential equations~\cite{Murray,Hutt+Atay_PhysicaD05}. To illustrate this relation, assume a one-dimensional
 spatial field $u(x)$ and the spatial kernel function $K(x)$. Then it is 
\begin{eqnarray*}
\int_{-\infty}^\infty K(x-y)u(y)dy=\int_{-\infty}^\infty\tilde{K}(k)\tilde{u}(k)e^{ikx}dk\label{eqn_int}
\end{eqnarray*} 
with the Fourier transforms $\tilde{K}$ and $\tilde{u}$ of $K$ and $u$, respectively. Then choosing 
 $K(x)=\exp(-|x|/\sqrt{D})/2\sqrt{D}$, it is $\tilde{K}(k)=1/(1+Dk^2)=1-Dk^2+D^2k^4+\cdots$ and Eq.~(\ref{eqn_int}) reads 
\begin{eqnarray*}
\int_{-\infty}^\infty K(x-y)u(y)dy&\approx& u(x,t)+D\Delta u(x,t)\nonumber\\
&&+D^2\Delta^2u(x,t) + \cdots
\end{eqnarray*} 
with $\Delta=\partial^2/\partial x^2$. This means that fast-decaying integral kernels $K(x)$ with
 $\sqrt{D}\ll 1$ represents a local diffusion process with diffusion constant $D$, 
while slowly-decaying integral kernels  yield higher orders of spatial derivatives and thus represent
long-ranged or nonlocal interactions. This procedure has been studied in some more detail in 
a previous work~\cite{Hutt+Atay_PhysicaD05} and it turned out that integral-differential equations
 generalize some types of partial differential equations. In this context the present work extends 
the previous work by the study of a more general spatio-temporal model. \\
In the research field of spatially extended systems, let the underlying model be partial differential 
equations or integral-differential equations, it is well-known that local excitation and lateral 
inhibition (local inhibition and lateral excitation) may yield stationary (non-stationary) spatial patterns in one or more spatial dimensions~\cite{Cross93,Hutt_Network03}. However, a recent study 
showed that local inhibition and lateral excitation may yield stationary instabilities as well 
for gamma-stributed spatial interactions~\cite{Hutt+Atay_PhysicaD05}. This 
finding let us look more carefully on the nature of spatial interactions and the classification 
criteria for these instabilities. The present work investigates wether there are traveling waves 
in local excitatory and lateral inhibitory systems, which may contrast to the findings in previous 
studies. \\
The work is structured as follows. The subsequent section introduces and motivates the generic 
model studied and presents the steps of corresponding linear stability analysis about a 
stationar state. Then section~\ref{sec_appl} applies the model to a field of harmonically damped
oscillators, which are coupled by nonlocal excitation and inhibition. The analytically 
results are confirmed by numerical simulations. The subsequent summary closes the work.

\section{The generic model}
The model studied presumes identical elements which obey the (non-)linear differential equation 
\begin{eqnarray*}
\hat{T}_tV(x,t)&=&h[V(x,t)]~.
\end{eqnarray*}
Here $V(x,t)$ represents the scalar activity variable, $\hat{T}_t=\hat{T}_t(\partial/\partial t)$ denotes
 the temporal operator and the functional $h[V(x,t)]$ defines the local dynamics at spatial location $x$.
Let the elements $\{x\}$ be located in a spatial domain and coupled to each other. This configuration 
reflects a topological network with identical elements
%see e.g.~\cite{Zhou+Kurths06}, 
in which no spatial metric is considered, or a topological network embedded in a physical space~\cite{Xulvi-Brunet+Sokolov02,Jost+Joy02}. In the the present model the single 
elements are connected according to a probability density function similar 
to topological networks but their connectivity function depend on the Euclidean distance between the
 elements similar to spatial networks. Further the spatial coupling is presumed homogeneous in space, 
that is the coupling is dependent on the spatial distance between the elements. These
properties describe a scale-free lattice~\cite{Yang04,Karmakar05}. Moreover, the presented 
model presumes infinitely dense elements, that is we consider a continuous field. This contrasts to
 networks whose major element is the discreteness. However in biological systems, e.g. in the brain, 
the single cells are dense and the continuum limit holds~\cite{vHemmen04}. 
Furthermore the speed by which the activity is transmitted between single elements shall be finite.
This finite propagation speed also contrasts to most spatial networks, which frequently
neglect propagation delay effects.

Considering all previous aspects, the element at spatial location $x$ obeys the integral-differential equation
\begin{eqnarray*}
\hat{T}_tV(x,t)&=&h[V(x,t)]+I(x,t)\nonumber\\
&&+\int_\Omega K(x-y)f\left[V(y,t-\frac{|x-y|}{v_K})\right]\\
&&+L(x-y)g\left[V(y,t-\frac{|x-y|}{v_L})\right]~dy\label{eqn_start}.
\end{eqnarray*}
Here $I(x,t)$ represents the external input and the kernel functions $K,~L$ represent the probability 
density functions of spatial connections in two 
networks, and  thus $K(x)$ and $L(x)$ are normalized to unity in the spatial domain $\Omega$. In 
the following we assume $\Omega={\mathcal R}$. The functionals $f[V]$ and $g[V]$ represent the 
(non-)linear coupling functions. Further the 
model considers propagation delays between spatial locations $x$ and $y$ at distance $|x-y|$ due to 
the finite transmission speed $v_K$ and $v_L$ corresponding to the two networks. This model allows the study of a wide range of combined spatial interaction. For instance 
spatial systems involving both mean-field and local interactions may be modeled by $K(x)=const$ 
and $L(x)=\delta(x)$, respectively. The subsequent paragraphs focus on the combination of excitatory 
and inhibitory interactions with kernel functions $K(x)$ and $L(x)$, respectively. In addition we 
specify $f[V]=a_KS[V],~g[V]=a_LS[V]$ with the nonlinear functional $S[V]$ and the total amount of 
excitation and inhibition $a_K$ and $a_L$, respectively. Further we presume $a_K>a_L$. 
For $h(V)=0$, Eq.~(\ref{eqn_start}) represents a well-studied nonlocal model of neural 
populations~\cite{Pinto01a,Coombes05_2,Hutt_Network03,Hutt+Atay_PhysicaD05,Hutt_PREBR04}.

\begin{figure}
\centerline{\includegraphics[width=6cm]{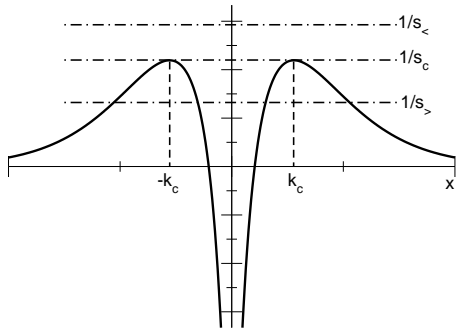}}
\caption{Sketch to illustrate the mechanism of non-stationary instabilities. The plot shows the right hand side of Eq.~(\ref{eqn_k}) 
(solid line) and its left hand side (dot-dashed line). If the stability threshold is not reached, i.e. $S^\prime=s_< < S^\prime_c$, the system is stable. In contrast, if $S^\prime=s_c=S^\prime_c$ ($S^\prime=s_> >S^\prime_c$) the system is marginal stable (unstable).
\label{fig_stab}}
\end{figure}

First let us determine the stationary state which is constant in space and time. Applying 
a constant external input $I_0$, we find the implicit equation $0=h(V_0)+f(V_0)+g(V_0)+I_0$. Here 
we used the fact that the kernel functions are normalized to unity. Then small deviations 
$u(x,t)$ about the stationary state obey a linear evolution equation and relax in time to the 
stationary state $V_0$ with $u(x,t)\sim e^{\lambda t},~\lambda\in~{\mathcal C}$ if $Re(\lambda)<0$. 
Then the subsequent expansion of $u(x,t)$ in the continuous Fourier basis $\{e^{ikz}\}$ yields 
the implicit condition 
\begin{eqnarray}
&&T(\lambda)=h_0+S^\prime\int_\Omega dy\,\nonumber\\
&&\times\left(a_KK(y)e^{-\lambda|y|/v_K}-a_LL(y)e^{-\lambda|y|/v_L}\right)e^{iky}\nonumber\\\label{eqn_lambda}
\end{eqnarray}
with $h_0=\partial h/\partial V$, $S^\prime=\partial S/\partial V$ computed at $V=V_0$. Since the 
external input drives the system, its stability is subjected to $I_0$ and subsequently $S^\prime$
 represents the reasonable control parameter. Equation~(\ref{eqn_lambda}) is difficult to solve 
exactly for $\lambda$. In a previous study a more specific model has been studied in detail for stationary 
and non-stationary instabilities~\cite{Hutt+Atay_PhysicaD05}. We follow its approach and consider 
large but finite propagation speeds. Then it is $\exp(-\lambda |z|/v)\approx 1-(\lambda /v)|z|$
for $|\lambda|/v_K,~|\lambda_L|/v_L\ll 1$ and (\ref{eqn_lambda}) reads 
\begin{eqnarray}
T(\lambda)\approx h_0+S^\prime\tilde{K}_0(k)-\lambda S^\prime\tilde{K}_1(k)~\label{eqn_L}.
\end{eqnarray}
with
\begin{eqnarray*}
\tilde{K}_0(k)&=&a_K\tilde{K}^{(0)}(k)-a_L\tilde{L}^{(0)}(k)\\
\tilde{K}_1(k)&&=\frac{a_K}{v_K}\tilde{K}^{(1)}(k)-\frac{a_L}{v_L}\tilde{L}^{(1)}(k)\\
\end{eqnarray*}
with $\tilde{F}^{(n)}(k)=\int_{-\infty}^\infty F(z)|z|^n\exp(-ikz)dz$. The term $\tilde{K}_0(k)$ and
 $\tilde{K}_1(k)$ represents the Fourier transform and the first kernel Fourier moment of
 $a_KK(x)+a_LL(x)$, respectively. Essentially let us specify the temporal 
evolution of a single element to a damped oscillator with 
\begin{eqnarray*}
\hat{T}_t=\frac{\partial^2}{\partial t^2}+\gamma\frac{\partial}{\partial t}+1 .
\end{eqnarray*} 
Then the stability threshold of stationary and non-stationary instabilities read 
\begin{eqnarray}
\frac{1}{S^\prime_c}&=&\frac{\tilde{K}_0(k_c)}{1-h_0}\quad\mbox{(stationary instability)}\label{eqn_k1}\\
\frac{1}{S^\prime_c}&=&-\frac{\tilde{K}_1(k)}{\gamma}\quad\mbox{(non-stationary instability)}\nonumber\\\label{eqn_k}
\end{eqnarray}
Hence the global maximum of $\tilde{K}_0(k_c)$ and $-\tilde{K}_1(k_c)$ define the critical 
wavenumber $k_c$ and the critical threshold $S^\prime_c$ of stationary and non-stationary instabilities, 
respectively. Figure~\ref{fig_stab} illustrates this result for non-stationary instabilities.
In the following, we call an emerging non-stationary phenomena a wave instability, if $k_c\ne 0$. 
and the instability with $k_c=0$ is called a global oscillation. Moreover it is interesting to note 
that the threshold of the non-stationary instability does not depend on the local 
(non-)linear dynamics of the elements $h(V)$. 

\section{Application to specific model }\label{sec_appl}
\begin{figure}
\includegraphics[width=8cm]{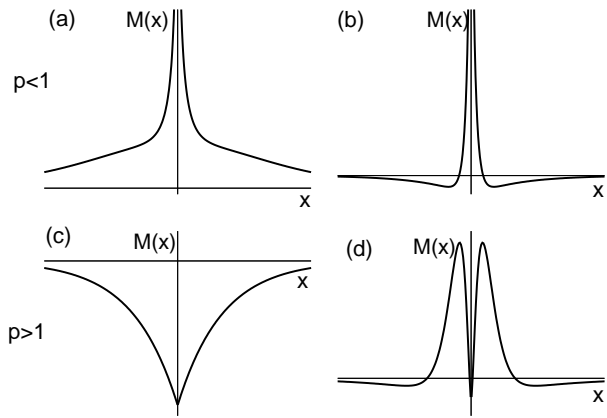}
\caption{Illustration for some spatial interaction types for different parameters 
$p, q$ and $r$. Panel (a) shows the kernel function for $p<1,~q>r$ that yields global excitations 
while panel (b) reveals local excitation-lateral inhibition for $p<1,~q<r$. In panel (c) the 
kernel function has been illustrated for $p>1,~q<r$, which represents global inhibition, while 
panel (d) shows local inhibition-lateral excitation for $p>1,~q>r$. It is $M(x)=K(x)-L(x)$\label{fig_kernels}}
\end{figure}

First let us discuss briefly various types of nonlocal interactions. We choose the excitatory 
and inhibitory spatial kernel to the gamma function and the decreasing exponential function, resp.,
with 
\begin{eqnarray}
K(z)=\frac{1}{2q^p\Gamma(p)}|z|^{p-1}e^{-|z|/q}~,~L(z)=\frac{1}{2r}e^{-|z|/r}\label{eqn_kernels}~
\end{eqnarray}
Since the Eq.~(\ref{eqn_k}) does not depend on $h[V]$, we set $h[V]=0$. In other words, the single elements in the field are assumed damped harmonic
oscillators, which are coupled by an excitatory and an inhibitory network.
The parameter $r$ defines the spatial range of the inhibitory interaction, while
both $p$ and $q$ represent the spatial constants of the excitatory nonlocal interaction. 
The scase $p=1$ is well-studied, see e.g.~\cite{Hutt_Network03}, and the combination of $q$ and $r$ 
yield the four important cases, namely the global excitation, the global inhibition, 
the local excitation-lateral inhibition and the local inhibition-lateral excitation. Further 
the choice $p>1$ yields local inhibitory interactions or local inhibitory-lateral 
excitatory interactions~\cite{Hutt+Atay_PhysicaD05}. The present work focus on the case $p<1$ 
which yields the local excitation or the local excitation-lateral inhibition. Please see Fig.~\ref{fig_kernels} for the corresponding illustration of the cases $p<1$ and $p>1$.
In order to gain the critical wavenumber of the wave instability, the kernel Fourier moment in Eq.~(\ref{eqn_k}) is computed to
\begin{eqnarray}
\tilde{K}^{(1)}(k)&=&\frac{pq\cos((p+1)\arctan(kq))}{(1+q^2k^2)^{(p+1)/2}}\label{eqn_K1e}\\
\tilde{L}^{(1)}(k)&=&\frac{r\cos(2\arctan(kr))}{(1+r^2k^2)}\label{eqn_K1i}
\end{eqnarray}
Since $\tilde{K}_1(k)\to 0$ for $|k|\to\infty$, there is a maximum of $-\tilde{K}_1(k)$ at some $|k|=|k_c|$ with $-\tilde{K}_1(k_c)>0$ if $-\tilde{K}_1(0)>0$ and $-d^2\tilde{K}_1(0)/dk^2>0$.
These sufficient conditions read
\begin{eqnarray}
\xi>p/a_r~,~\xi^3<\frac{v_r}{6a_r}p(p+1)(p+2)\label{eqn_suffi}
\end{eqnarray}
with the new parameters $\xi=r/q,~v_r=v_L/v_K$ and $a_r=a_L/a_K<1$. 
\begin{figure}
\includegraphics[width=8cm]{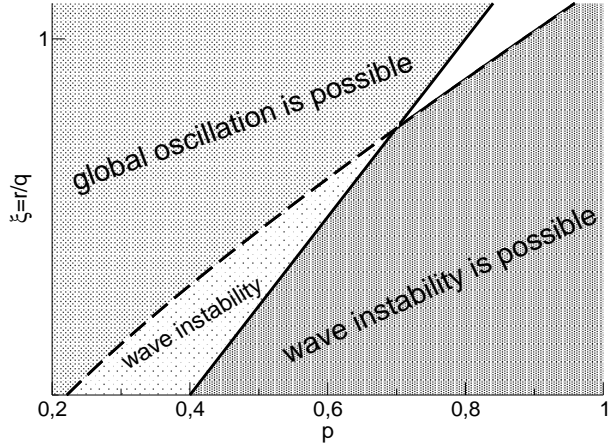}
\caption{The occurrence of oscillatory instabilities subject to the parameters $\xi=r/q$ and $p$. The dashed and thin solid line 
represent the function $\xi=(v_rp(p+1)(p+2)/6a_r)^{1/3}$ and $\xi=p/a_r$, respectively, taken 
from Eq.~(\ref{eqn_suffi}). Here it is $v_r=1,~a_r=0.8$~.\label{fig_xip}}
\end{figure}
Figure~\ref{fig_xip} illustrates the different parameter regimes of the conditions. At first 
it turns out that there is large parameter regime which allows for non-stationary instabilities. 
In order to gain 
further insight to the nature of spatial interactions for $p<1$, Fig.~\ref{fig_xipmultiple} shows the parameter regimes and the different number of roots of the kernel 
function $K(z)$. We find a parameter regime of single roots, which reflects local 
excitation-lateral inhibition interaction, and a parameter regime of no roots, i.e. 
global excitation. Moreover, there is a parameter regime of multiple roots which, for instance, 
reflects local excitation - mid-range inhibition - lateral excitation and allows for wave 
instabilities. \\
\begin{figure}
\includegraphics[width=8cm]{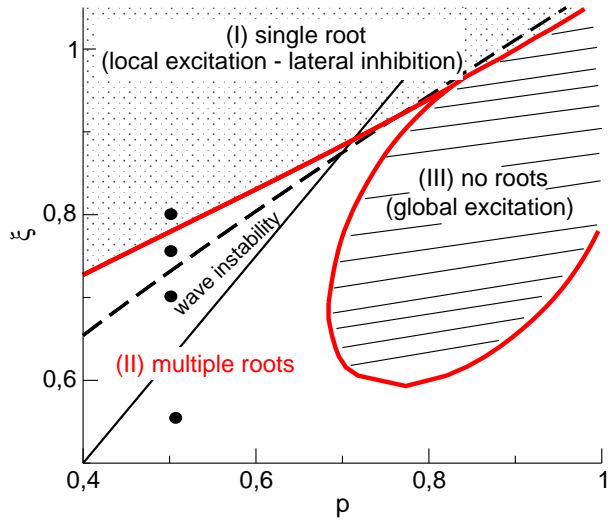}
\caption{Number of zero-croosings, i.e. roots, of the kernel function $K(z)$ and the corresponding nature of spatial interactions subject to the parameters $\xi$ and $p$. The dashed and thin solid line are 
taken from Fig.~\ref{fig_xip}. The red lines delimits the novel parameter regime of multiple roots.
In addition to the parameters from Fig.~\ref{fig_xip}, it is $a_K=10,~a_L=8, v_K=v_L=1$ and $q=1$.
\label{fig_xipmultiple}}
\end{figure}
To illustrate the relation of the spatial interaction type and the resulting non-stationary 
instability,
Figure~\ref{fig_kernel+kernelfourier} plots the kernel functions $K(z)$ and the corresponding negative 
first kernel Fourier moments $-\tilde{K}_1(k)$ for four different cases of 
spatial interactions. These specific cases are denoted by filled dots in Fig.~\ref{fig_xipmultiple}. We 
find local excitation-lateral inhibition for $\xi=0.8$ (top row, left panel in Fig.~\ref{fig_xipmultiple}). 
Since the global maximum of $-\tilde{K}_1(k)$ gives the critical wavenumber $k_c$, this case yields global 
 oscillations with $k_c=0$ (top row, right panel in Fig.~\ref{fig_xipmultiple}). For $\xi=0.75$, now the 
kernel function exhibits two roots which still leads to global oscillations. 
In case of $\xi=0.70$ the sufficient condition for wave instabilities are fulfilled, as 
$-\tilde{K}_1(k)$ reveals a critical wavenumber $k_c\ne 0$, see Fig.~\ref{fig_xipmultiple}. 
Moreover, the spatial interaction function reveals local excitation, mid-range inhibition and 
lateral excitation, which has not been found yet in previous studies. At last for $\xi=0.55$ there is also a wave instability showing local excitation, mid-range inhibition and lateral excitation.
\begin{figure}
\includegraphics[width=8.5cm]{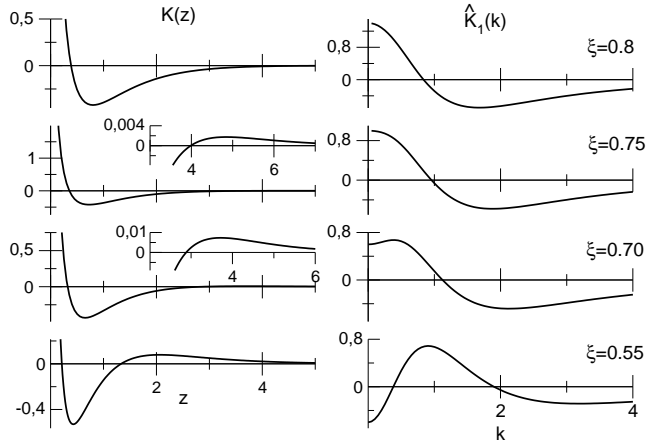}
\caption{The kernel functions $a_KK(z)+a_LL(z)$ and the negative first kernel Fourier moment $\tilde{K}_1(k)$ for the four different cases
taken from Fig.~\ref{fig_xipmultiple}. The insets in the left coloumn represent foci of the corresponding plot to illustrate
the additional lateral excitation. The parameters are taken from Fig.~\ref{fig_xipmultiple} and 
the different values of $\xi$ correspond to the different dots in Fig.~\ref{fig_xipmultiple}.
\label{fig_kernel+kernelfourier}}
\end{figure}
In addition to this study, Fig.~\ref{fig_kernel+kernelfourier2} compares the spatial interaction 
function, the Fourier transform $\tilde{K}_0$ and $-\tilde{K}_1(k)$ for two values of $\xi$. This comparison 
is necessary as the stationary instability may occur first before the wave instability while 
increasing the control parameter from small values. For both parameter values of $\xi$ the global 
maximum of $-\tilde{K}_1(k)$ exceeds the global maximum of $\tilde{K}_0$ and thus a wave 
instability and not a stationary instability may occur. Thus Fig.~\ref{fig_kernel+kernelfourier2} 
supports the previous findings.\\
\begin{figure}
\includegraphics[width=8.5cm]{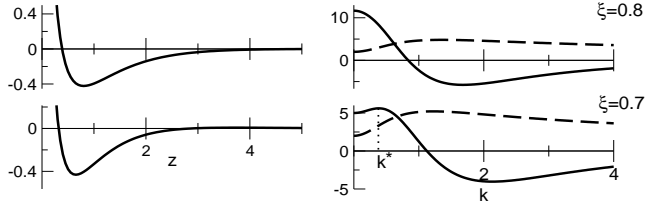}
\caption{The comparison of the kernel functions $a_KK(z)+a_LL(z)$ (left coloumn), the Fourier 
transform $\tilde{K}_0(k)$ (dashed line in right coloumn) and 
$-\tilde{K}_1(k)$ (solid line in right coloumn) for 
two values of $\xi$. Additional parameters are taken from Fig.~\ref{fig_xipmultiple}.
\label{fig_kernel+kernelfourier2}}
\end{figure}
To visualize and confirm the analytical findings, finally we integrate the integral-differential equation
~(\ref{eqn_start}) numerically. The Euler integration scheme is applied for the time evolution 
integration and the Monte-Carlo method VEGAS~~\cite{VEGAS,GSL} computed the spatial integration. 
This specific Monte-Carlo integration method combines stratified and importance sampling, which 
is important to gain good estimates of the divergent excitatory kernel function. Moreover 
boundary conditions are set periodic with the period $L$ while the initial 
conditions have to be defined in the time interval $t_0\in[-L/v;0]$.  We have chosen $V(z,t_0)=\cos(k_cx)+0.5*\cos(5k_cx)$.
\begin{figure}
\includegraphics[width=9cm]{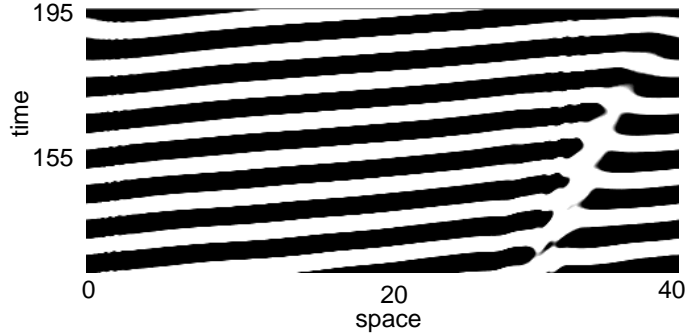}
\caption{Space-time plot of the integrated activity for $p=0.5,~r=0.7$ shows the wave instability 
for local excitation - lateral inhibition interactions. Additional parameters are
$a_K=10,~a_L=8,~v_K=v_L=0.6,~\gamma=0.2,~I_0=2.8,~dx=0.089,~dt=0.08$ and $h[V]=0,~S[V]=1/(1+\exp(-1.84(V-3.0)))$. These parameters yields $k_c=0.4$.
\label{fig_wave}}
\end{figure}
Figure~\ref{fig_wave} shows the wave instability for $p=0.5,~\xi=0.7$  discussed above, cf. Fig.~\ref{fig_xipmultiple}, and confirms the analytical finding.

\section{Summary}
The present work describes the nonlocal spatial interaction of single elements in a continuous 
field. The application to damped oscillators, which are coupled by both nonlocal excitation and 
inhibition, reveals a new mechanism of wave instabilities. In contrast to previous 
findings in pattern forming systems, we  find wave instabilities for local excitation and 
lateral inhibition. To obtain more information on the corresponding spatial interaction,
we investigated additionally the spatial interaction function. It turns out that 
various spatial interaction types exhibit wave instabilities, as local excitation-lateral 
inhibition and local excitation-midrange inhibition-lateral excitation. Hence 
the sign of the spatial interaction function does not carry the full information
of the expected instability patterns in spatial systems with finite propagation delay. We have 
observed that the comparison of the global maximum of the Fourier component and global maximum 
of the negative first kernel Fourier moment allows for a reliable estimation of the 
instability type.\\
Future work shall study the classifiers in vectorial models, as the nonlocal version 
of the Bruesselator, and the classification scheme in two- and three-dimensional systems.

\begin{appendix}
\section*{Appendix: The kernel Fourier moments}
The n-th kernel Fourier moment of the specific kernels (\ref{eqn_kernels}) can be computed 
from 
\begin{eqnarray*}
 I(p)&=&\int_0^\infty x^{p-1}e^{-z/q}\cos(kx)\,dx\nonumber\\
&=&\Gamma(p)\left(\frac{1}{q^2}+k^2\right)^{-p/2}\cos\left(p\arctan(kq)\right)\label{Fourier}
\end{eqnarray*}
Then we find for the excitatory kernel 
\begin{eqnarray*}
 \tilde{K}^{(n)}(k)&=&\frac{1}{q^p\Gamma(p)}~I(p+n)\\
&=&\frac{\Gamma(p+n)}{\Gamma(p)}\frac{q^n\cos\left((p+n)\arctan(kq)\right)}{(1+q^2k^2)^{(p+n)/2}}~.
\end{eqnarray*}
In addition, we find
\begin{eqnarray*}
-\frac{\partial^2}{\partial k^2}\tilde{K}^{(n)}(k)= \tilde{K}^{(n+2)}(k)~.
\end{eqnarray*}
Since the inhibitory kernel is a specific case of the excitatory kernel with $p=1$ and $q=r$, we 
obtain Eqs.~(\ref{eqn_K1e}), (\ref{eqn_K1i}) from the previous expressions.

\end{appendix}

%\bibliographystyle{plain}
%\bibliography{paper}% Produces the bibliography via BibTeX.

\end{document}